\documentclass[aps, prb, twocolumn, floatfix, superscriptaddress, nofootinbib]{revtex4-2}
\usepackage{amsmath, amssymb, graphicx, epsfig, xcolor}
\usepackage[colorlinks=true, allcolors=blue]{hyperref}

\begin{document}

\title {Prediction of ambient pressure superconductivity in cubic ternary hydrides with MH$_6$ octahedra}

\author{Feng Zheng$^{\dag}$}
\address{School of Science, Jimei University, Xiamen 361021, China}
\address{Department of Physics, Xiamen University, Xiamen 361005, China}

\author{Zhen Zhang$^{\dag}$}
\address{Department of Physics and Astronomy, Iowa State University, Ames, IA 50011, USA}

\author{Shunqing Wu$^{\ast}$}
\address{Department of Physics, Xiamen University, Xiamen 361005, China}

\author{Qiubao Lin}
\address{School of Science, Jimei University, Xiamen 361021, China}

\author{Renhai Wang}
\address{School of Physics and Optoelectronic Engineering, Guangdong University of Technology, Guangzhou 510006, China}

\author{Yimei Fang}
\address{Department of Physics, Xiamen University, Xiamen 361005, China}

\author{Cai-Zhuang Wang}
\address{Department of Physics and Astronomy, Iowa State University, Ames, IA 50011, USA}
\address{Ames National Laboratory, U.S. Department of Energy, Ames, IA 50011, USA}

\author{Vladimir Antropov}
\address{Department of Physics and Astronomy, Iowa State University, Ames, IA 50011, USA}
\address{Ames National Laboratory, U.S. Department of Energy, Ames, IA 50011, USA}

\author{Yang Sun$^{\ast}$}
\address{Department of Physics, Xiamen University, Xiamen 361005, China}

\author{Kai-Ming Ho}
\address{Department of Physics and Astronomy, Iowa State University, Ames, IA 50011, USA}

\def\thefootnote{$\dag$}\footnotetext{These authors contributed equally to this work.}\def\thefootnote{\arabic{footnote}}
\def\thefootnote{$\ast$}\footnotetext{Email: wsq@xmu.edu.cn (S.W.); yangsun@xmu.edu.cn (Y.S.)}\def\thefootnote{\arabic{footnote}}

\date{Jan. 28, 2024}

\begin{abstract}
Exploring high-temperature superconducting (high-$T_c$) material at ambient pressure holds immense significance for physics, chemistry, and materials science. In this study, we perform a high-throughput screening of strong electron-phonon interactions in X$_2$MH$_6$ compounds (X = Li, Na, Mg, Al, K, Ca, Ga, Rb, Sr, and In; M are $3d$, $4d$, and $5d$ transition metals). These compounds have a cubic structure featuring an MH$_6$ octahedron motif. Our screening calculations suggest that 26 compounds exhibit dynamic stability and strong electron-phonon coupling. Among these 26 compounds, Mg$_2$RhH$_6$, Mg$_2$IrH$_6$, Al$_2$MnH$_6$, and Li$_2$CuH$_6$ show promising energetic stability and $T_c$ of more than 50 K at ambient pressure. This study underscores promising high-$T_c$ compounds at ambient pressure with distinctive MH$_6$ motifs.
\end{abstract}

\maketitle

\section{Introduction}
The idea of metallic hydrogen processing a significant high-pressure superconductivity was suggested by Abrikosov \cite{1} and Ashcroft \cite{2} a long time ago. However, the problem of making metallic hydrogen remains unsolved \cite{3} and the search for corresponding superconductivity was shifted to the area of systems with partial hydrogen content, assuming that in such compounds, one can reduce the physical pressure of hydrogen metallization due to chemical pressure. In this way, the first superconducting hydride Th$_4$H$_{15}$ with $T_c$ $\sim$ 8.05-8.35 K was reported in 1970 \cite{4}. After that, the superconductivity of the Pd-H system was discovered with a $T_c$ of 8-9 K \cite{5}. The Pd-(Cu, Ag, and Au)-H system was also deeply studied with $T_c$ of 16.6, 15.6, and 13.6 K \cite{6}. It appears that chemical pressure was not strong enough, and hydride superconductivity research in recent years has shifted to using both physical high-pressure and chemical pressure ideas with many successful discoveries, including theoretical ones. Many H-rich compounds \cite{7,8,9,10,11,12,13,14,15,16} have been predicted to be promising conventional superconductors with high $T_c$ under high pressure, using state-of-the-art crystal structure search methods \cite{17,18,19,20,21}. Among them, H$_3$S \cite{22}, LaH$_{10}$ \cite{23, 24}, CaH$_6$ \cite{25, 26}, YH$_9$ \cite{27,28,29} and YH$_6$ \cite{27} have been experimentally confirmed to have a $T_c$ greater than 200 K under high pressures. These remarkable discoveries represent a significant milestone, suggesting that conventional mechanisms through electron-phonon interaction can achieve superconductivity close to room temperature \cite{30, 31}. Although hydrides can trap H atoms within their host lattice to achieve superconductivity, the extremely high-pressure conditions required to stabilize the structures is still challenging. It has only been attained in a few experiments and hinders their practical applications. Therefore, the research on superconductivity in hydrides necessitates a shift toward exploring novel H-rich compounds that can be superconducting at ambient pressure.

Extensive theoretical calculations have been conducted to shed light on ambient-pressure superconducting hydrides. Vocaturo \textit{et al.} predicted that PdCuH$_2$ possesses superconducting properties with $T_c$ $\sim$ 34 K at ambient pressure \cite{32}. Al$_4$H ($Pm\bar{3}m$ space group) \cite{33} and (Be$_4$)$_2$H (by combining $hcp$ H and Be) \cite{34} were also predicted to have $T_c$ of 54 and 72-84 K, respectively. However, these two structures are both metastable with energy above convex hull ($E_d$) of 136 and 202 meV/atom \cite{35}. Using machine-learning accelerated high-throughput techniques, Tiago \textit{et al.} identified several hydride and hydrogen-containing materials with high-$T_c$ at ambient pressure, such as Li$_2$PdH$_2$ (43 K), PdH (37 K), ZrH$_3$ (24.9 K), and KCdH$_3$ (23.4 K) \cite{36}. Very recently, by employing \textit{ab initio} random structure search (AIRSS) across the periodic table, Dolui \textit{et al.} predicted cubic Mg$_2$IrH$_6$ to be a superconductor at ambient pressure, with the $T_c$ up to 160 K \cite{37}. The Mg$_2$IrH$_6$ is isostructural with the Mg$_2$FeH$_6$ family \cite{38}, which was previously synthesized. Sanna \textit{et al.} also investigated the superconducting properties of the Mg$_2$MH$_6$ family (M = Rh, Ir, Pd, and Pt) and suggested their superconducting $T_c$ are in the range of 45-80 K \cite{35}. The crystal structure of Mg$_2$MH$_6$ is a cubic phase with an MH$_6$ octahedron structural motif. A natural question remains whether other ternary superconducting hydrides can be stabilized in this structural framework at ambient pressure.

In this paper, we employ our recently developed fast screening method \cite{39} to investigate the possible superconductivity in cubic X$_2$MH$_6$ compounds (X = Li, Na, Mg, Al, K, Ca, Ga, Rb, Sr and In; M is $3d$, $4d$, and $5d$ transition metal) at ambient pressure. The screening is based on the electron-phonon coupling strength (EPC) at the $\Gamma$ point ($\lambda_\Gamma$), which was found to be an efficient descriptor to identify conventional superconducting materials in hydrides and borides \cite{14, 39,40,41,42,43,61}. We will explore their superconducting behavior, energetic and dynamic stabilities under ambient conditions.
\\

\section{Computational methods}
The structure optimizations were performed by using the projector-augmented wave (PAW) \cite{44} representations with density functional theory implemented in the Vienna \textit{ab initio} simulation package (VASP) \cite{45, 46}. The exchange and correlation energy is treated within the spin-polarized generalized gradient approximation (GGA) and parameterized by Perdew-Burke-Ernzerhof (PBE) formula \cite{47}. A plane-wave basis was used with a kinetic energy cutoff of 520 eV. Brillouin zone integrations were approximated by using special \textbf{k}-point sampling of the Monkhorst-Pack scheme \cite{48} with $2\pi\times0.02$ $\rm{\AA}^{-1}$. Lattice vectors and atomic coordinates were fully relaxed until the force on each atom was less than 0.01 eV$\cdot\rm{\AA}^{-1}$. The fast screening of EPC constant $\lambda_\Gamma$ at the Brillouin zone center was carried out based on the frozen-phonon method \cite{39}. Here, the $\lambda_\Gamma$ can be defined by
\begin{equation}
\lambda_{\Gamma}=\sum_\nu \lambda_{\Gamma \nu},
\end{equation}
where $\sum_\nu$ indicates the summation of all modes at zone-center $\Gamma$. $\lambda_{\Gamma \nu}$ is defined by
\begin{equation}
\lambda_{\Gamma \nu}=\frac{\widetilde{\omega}_{\Gamma \nu}^2-\omega_{\Gamma \nu}^2}{4 \omega_{\Gamma \nu}^2}
\end{equation}
where the $\omega_{\Gamma \nu}$ and $\widetilde{\omega}_{\Gamma \nu}$ are screened and unscreened phonon frequencies of mode $\nu$ at zone-center, respectively. The zone-center phonon was computed by the PHONOPY software \cite{49, 50}, with a \textbf{k}-point sampling grid of $2\pi\times0.02$ $\rm{\AA}^{-1}$ spacing and a criterion of self-consistent calculation 10$^{-8}$ eV.

The full Brillouin-zone EPC calculation was performed with the Quantum ESPRESSO (QE) code \cite{51, 52} based on the density-functional perturbation theory (DFPT) \cite{53}. The pseudopotentials (stringent norm-conserving set) from the PSEUDODOJO project \cite{54} for PBE functional were used. The kinetic energy cutoffs were 120 Ry for wave functions and 600 Ry for potentials. Monkhorst-Pack's sampling scheme \cite{48} was adopted for Brillouin-zone sampling with a \textbf{k}-point grid of $2\pi\times0.02$ $\rm{\AA}^{-1}$. The DFPT calculations were performed with the \textbf{k}-mesh of exactly the sampling scheme and set the \textbf{q}-mesh to half of the \textbf{k}-mesh. The convergence threshold for self-consistency was $1 \times 10^{-14}$ Ry.

The calculations of superconducting $T_c$ are based on the Eliashberg spectral equation $\alpha^2 F(\omega)$ defined commonly now as
\begin{equation}
\alpha^2 F(\omega)=\frac{1}{2 \pi N\left(E_f\right)} \sum_{q \nu} \frac{\gamma_{q \nu}}{\hbar \omega_{q \nu}} \delta\left(\omega-\omega_{q \nu}\right),
\end{equation}
where $N(E_f)$ is the density of states at the Fermi level,  $\omega_{q \nu}$ denotes the phonon frequency of the mode $\nu$ with wave vector \textbf{q}. $\gamma_{q \nu}$ is the phonon linewidth defined as
\begin{equation}
\gamma_{q \nu}=\frac{2 \pi \omega_{q \nu}}{\Omega_{B Z}} \sum_{i j} \int d^3 k\left|g_{k, q \nu}^{i j}\right|^2 \delta\left(\epsilon_{q, i}-E_f\right) \delta\left(\epsilon_{k+q, j}-E_f\right),
\end{equation}
where $\epsilon_{q, i}$ and $\epsilon_{k+q, j}$ are eigenvalues of Kohn-Sham orbitals at given bands and vectors. \textbf{q} and \textbf{k} are wave vectors, and $i$ and $j$ denote indices of energy bands. $g_{k, q \nu}^{i j}$ is the EPC matrix element. The EPC constant $\lambda$ can be determined through summation over the first Brillouin zone or integration of the spectral function in frequency space,
\begin{equation}
\lambda=\sum_{q \nu} \lambda_{q \nu}=2 \int \frac{\alpha^2 F(\omega)}{\omega} d \omega,
\end{equation}
where the EPC constant $\lambda_{q \nu}$ for mode $\nu$ at wave vector \textbf{q} using Eq. (5) can be written as
\begin{equation}
\lambda_{q \nu}=\frac{\gamma_{q \nu}}{\pi \hbar N\left(E_f\right) \omega_{q \nu}^2} .
\end{equation}
The superconducting $T_c$ is determined with the analytical McMillan equation modified by the Allen-Dynes (A-D) equation \cite{55, 56},
\begin{equation}
T_c=\frac{\omega_{\log }}{1.2} \exp \left[\frac{-1.04(1+\lambda)}{\lambda\left(1-0.62 \mu^*\right)-\mu^*}\right],
\end{equation}
where $\omega_{\log }$ is the logarithmic average frequency
\begin{equation}
\omega_{\log }=\exp \left[\frac{2}{\lambda} \int \frac{\mathrm{d} \omega}{\omega} \alpha^2(\omega) F(\omega) \log \omega\right]
\end{equation}
and the effective screened Coulomb repulsion constant $\mu^*$ was 0.1 in our calculations. \\

\section{Results and Discussion}
\subsection{High throughput screening of superconductivity in X$_2$MH$_6$}

\begin{figure*}[t]
	\includegraphics[width=0.7\linewidth]{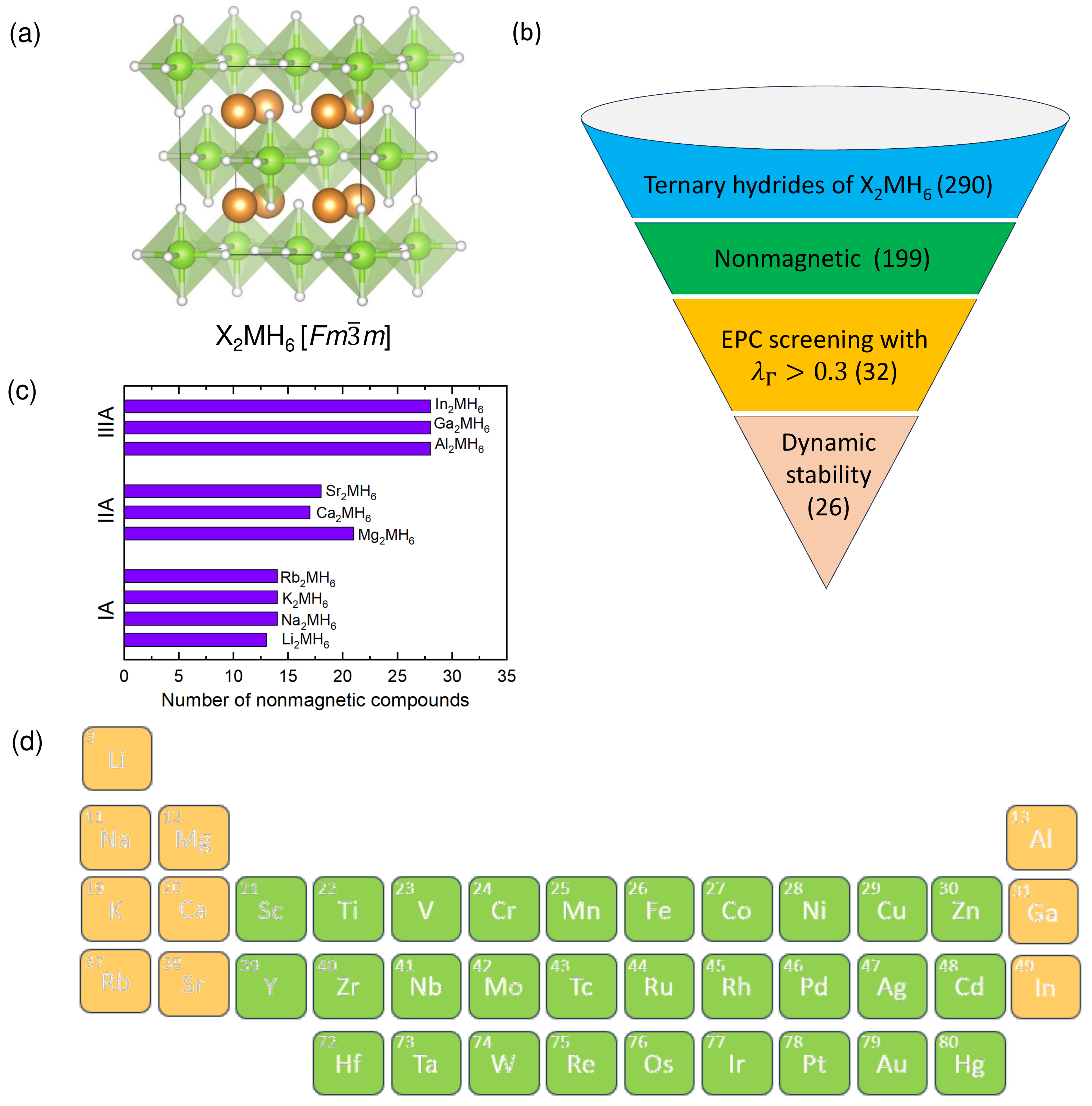}
	\caption{(a) The prototype structure of X$_2$MH$_6$. The X, M, and H atoms are colored in orange, green and white, respectively. (b) Schematic of screening workflow. (c) The number of nonmagnetic compounds of X$_2$MH$_6$ with X in different groups. (d) The elements involved in the screening. Orange indicates the elements of X. Green indicates the elements of M.}
\end{figure*}

Figure 1 summarizes our screening process of superconducting ternary hydrides at ambient pressure. As shown in Fig. 1(a), the prototype structure X$_2$MH$_6$ is a cubic phase with MH$_6$ octahedron structural motif, which is isostructural with Mg$_2$FeH$_6$ \cite{38}. By replacing X with Li, Na, Mg, Al, K, Ca, Ga, Rb, Sr, and In, and M with $3d$, $4d$, and $5d$ transition metals (seen Fig. 1d), we generate 290 candidates for screening. The screening criteria are shown in Fig. 1 (b). The X$_2$MH$_6$ compounds are all relaxed by the structural optimization at 0 GPa. After spin-polarized calculations, we remove magnetic compounds (magnetic moments are larger than 0.3 $\mu_B$/M) since the conventional superconductivity is mutually exclusive with magnetism. This step reduces the population of the candidate pool to 199 compounds. We find that the number of nonmagnetic compounds of X$_2$MH$_6$ can be influenced by the different groups of X ions, as shown in Fig. 1(c). X$_2$MH$_6$ with X in the IIIA group has the highest number of nonmagnetic compounds, followed by those in the IIA and IA groups. These nonmagnetic compounds are screened by the zone-center EPC strength calculations. By setting a threshold of 0.3 ($\lambda_{\Gamma}>0.3$), 32 materials are selected for further calculations. Among them, 26 compounds are dynamically stable. To evaluate their thermodynamic stability, the energy above the convex hull ($E_d$) is computed and shown in Table 1. The reference phases in the convex hulls are obtained from the Material Project database \cite{57}. The $E_d$ represents the energy of a material to decompose into a set of stable phases. If the $E_d$ = 0, it indicates that the phase is stable. However, a metastable phase with small $E_d$ can also be synthesized, given that many registered structures in the Inorganic Crystal Structure Database (ICSD) are metastable, according to the recent survey \cite{58}. We employ the $E_d$ less than 80 meV/atom as the criterion, corresponding to a typical thermal energy of $\sim$ 900 K \cite{59, 60}. After these screening, six low-energy superconductors are ultimately identified, i.e., Mg$_2$CoH$_6$, Mg$_2$RhH$_6$, Mg$_2$IrH$_6$, Ca$_2$PtH$_6$, Al$_2$MnH$_6$ and Li$_2$CuH$_6$. We then conduct full Brillouin-zone EPC and $T_c$ calculations and identify four compounds with $T_c$ greater than 50 K at ambient pressure, i.e., Mg$_2$RhH$_6$, Mg$_2$IrH$_6$, Al$_2$MnH$_6$ and Li$_2$CuH$_6$. The superconducting behaviors of these compounds are studied in the next section. Superconductivity calculations for the remaining compounds are listed in Supporting Information.

\begin{table}[t]
	\caption{The zone-center EPC strength ($\lambda_\Gamma$) and energy above convex hull ($E_d$) for 26 dynamically stable ternary hydrides.}
	\begin{ruledtabular}
		\begin{tabular}{lll}
			Ternary hydrides & $\lambda_\Gamma$ & $E_d$ (meV/atom)\\
			Mg$_2$RhH$_6$	& 6.23 & 0
\\
			Mg$_2$IrH$_6$	& 3.27 & 0
\\
			Ca$_2$PtH$_6$	& 0.75 & 9
\\
			Mg$_2$CoH$_6$	& 0.42 & 29
\\
			Al$_2$MnH$_6$	& 0.96 & 80
\\
			Li$_2$CuH$_6$	& 0.41 & 80
\\
			Ca$_2$PdH$_6$	& 0.40 & 97
\\
			Sr$_2$CuH$_6$	& 0.31 & 99
\\
			Ca$_2$CuH$_6$	& 0.32 & 109
\\
			Na$_2$CuH$_6$	& 0.30 & 112
\\
			Na$_2$AuH$_6$	& 0.31 & 113
\\
			Rb$_2$CuH$_6$	& 0.30 & 113
\\
			Al$_2$TcH$_6$	& 1.67 & 127
\\
			In$_2$TcH$_6$	& 1.27 & 151
\\
			Ga$_2$TcH$_6$	& 0.70 & 172
\\
			Mg$_2$PtH$_6$	& 1.56 & 185
\\
			In$_2$MnH$_6$	& 0.32 & 210
\\
			Li$_2$AuH$_6$	& 0.64 & 217
\\
			K$_2$AgH$_6$	& 0.30 & 259
\\
			Rb$_2$AgH$_6$	& 0.31 & 266
\\
			Na$_2$AgH$_6$	& 0.31 & 286
\\
			Rb$_2$CdH$_6$	& 0.35 & 324
\\
			Ga$_2$RuH$_6$	& 0.65 & 327
\\
			K$_2$CdH$_6$	& 0.65 & 386
\\
			Rb$_2$ZnH$_6$	& 1.35 & 422
\\
			Ga$_2$MnH$_6$	& 1.14 & 778
\\
		\end{tabular}
	\end{ruledtabular}
	\label{tab1}
\end{table}

\subsection{Superconductivity of Mg$_2$RhH$_6$, Mg$_2$IrH$_6$, Al$_2$MnH$_6$ and Li$_2$CuH$_6$}
\subsubsection{Mg$_2$RhH$_6$ and Mg$_2$IrH$_6$}

\begin{figure}[t]
	\includegraphics[width=1\linewidth]{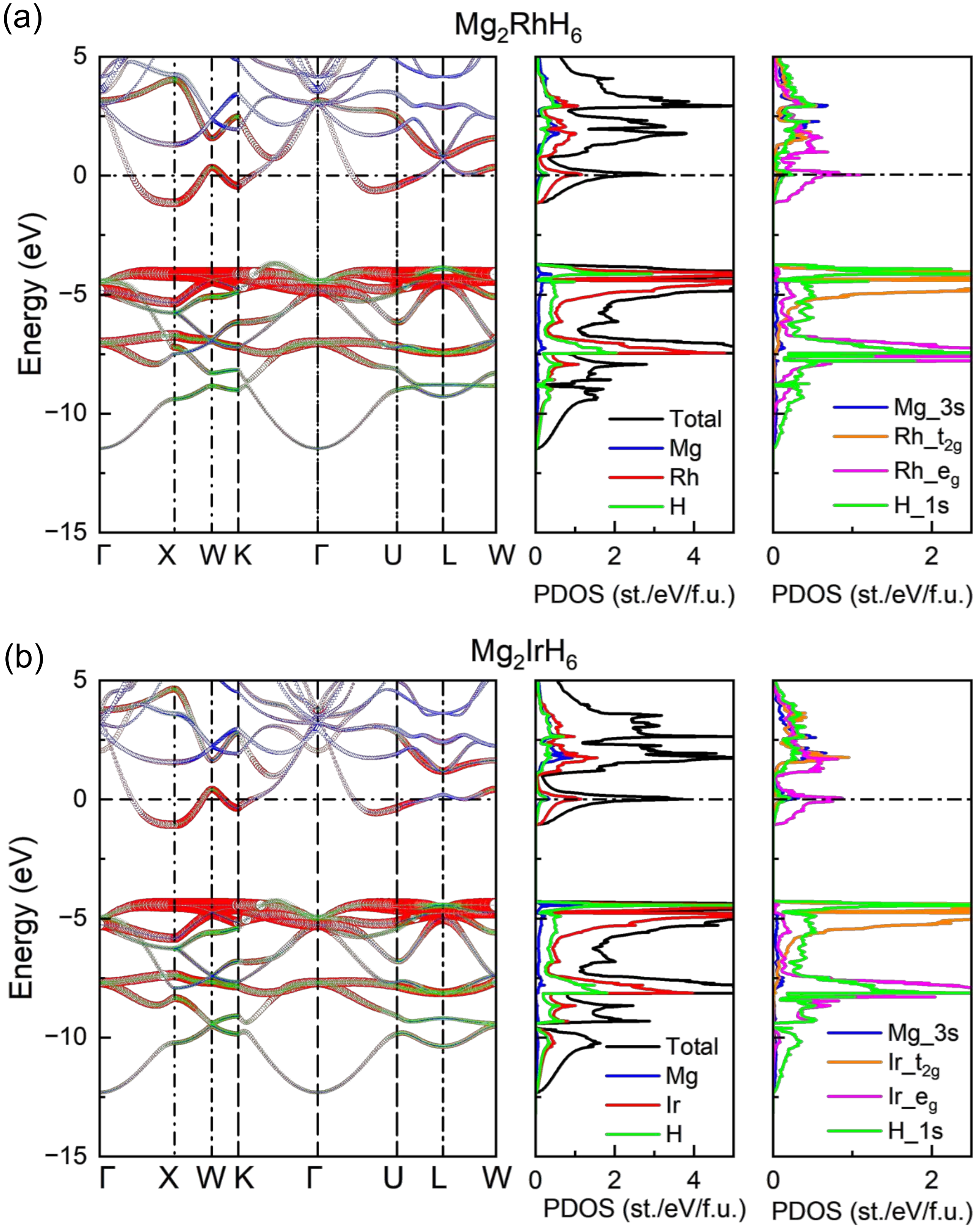}
	\caption{The electronic band structure and density of states projected into atoms and orbitals for (a) Mg$_2$RhH$_6$ and (b) Mg$_2$IrH$_6$.}
\end{figure}

Figure 2 shows the band structures and density of states projected into atoms and orbitals for Mg$_2$RhH$_6$ and Mg$_2$IrH$_6$. One can see that they have nearly identical electronic structures, which are both metallic with a van Hove singularity at the Fermi level ($E_f$). The calculated total DOS at $E_f$ are 3.09 and 3.17 eV$^{-1}$ per f.u. for Mg$_2$RhH$_6$ and Mg$_2$IrH$_6$, respectively. Near the $E_f$, the states are predominantly composed of transition metals $e_g^*$ antibonding states mixed with H-1$s$ ($\sim$ 0.30 and $\sim$ 0.28 eV$^{-1}$ per f.u. for Mg$_2$RhH$_6$ and Mg$_2$IrH$_6$, respectively) and Mg-3$s$ states. In the valence band of these two compounds, sharp peaks are observed in the $t_{2g}$ states, corresponding to the flat bands along the high-symmetry path. The $t_{2g}$ nonbonding bands of Mg$_2$RhH$_6$ and Mg$_2$IrH$_6$ exhibit a significant energy gap with their $e_g^*$ antibonding states of $\sim$ 2.6 and $\sim$ 3.3 eV, respectively. The $e_g$ bonding bands at a lower energy level also display a sharp peak.  We note that the summation of the projected DOS in the PAW sphere is smaller than the total DOS because of significant charge accumulations in the interstitial region. The intersititial contribution to the density of states at the Fermi level is significant in these systems.

\begin{figure}[t]
	\includegraphics[width=1\linewidth]{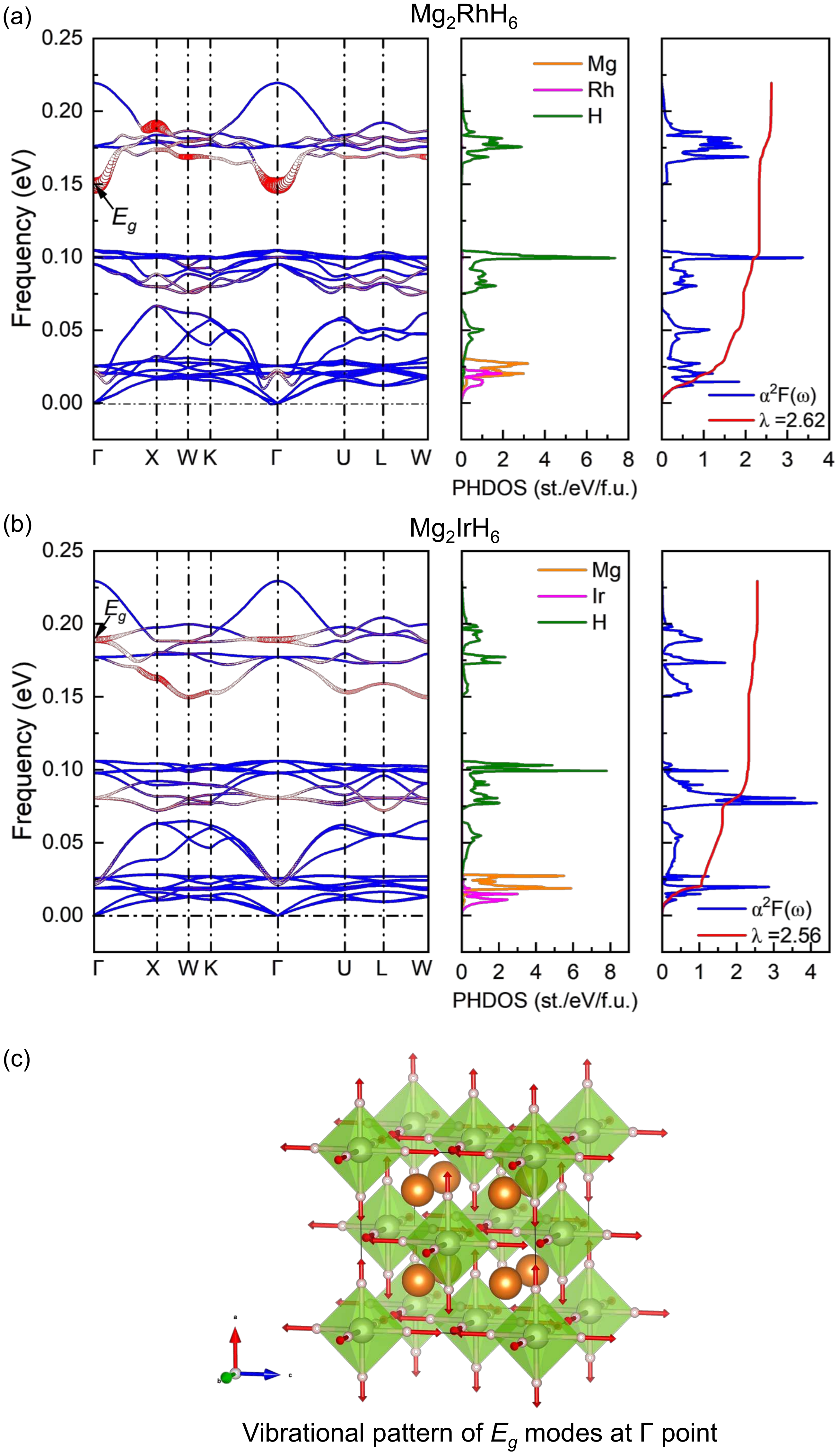}
	\caption{The $\gamma_{q \nu}$-weighted (red circles) phonon spectrum, projected phonon density of states (PHDOS) and Eliashberg spectral function $\alpha^2 F(\omega)$ for (a) Mg$_2$RhH$_6$ and (b) Mg$_2$IrH$_6$. (c) The vibration pattern for double degenerate $E_g$ modes at the $\Gamma$ point. The displacements are denoted by red arrows.}
\end{figure}

Figure 3 displays the phonon linewidth ($\lambda_{q \nu}$)-weighted phonon spectrum, the projected phonon density of states (PHDOS), and Eliashberg spectral function $\alpha^2 F(\omega)$ for Mg$_2$RhH$_6$ and Mg$_2$IrH$_6$. The phonon spectra of these two compounds are divided into three distinct regions. In the low-frequency region, the spectra are primarily dominated by transition metals and Mg atoms. The intermediate and high-frequency regions mainly arise from the vibrations of H atoms. As depicted in Fig. 3(a), for Mg$_2$RhH$_6$, the high-frequency H modes significantly contribute to the phonon linewidth, especially the double degenerate phonon modes $E_g$ at the $\Gamma$ point. Their vibrational modes are shown in Fig. 3(c), which involve the breathing vibrations of H atoms in the MH$_6$ octahedra. The intermediate-frequency region from $\sim$ 0.075 to $\sim$ 0.105 eV and low-frequency modes at $\sim$ 0.021 eV along $\Gamma$ point corresponding to H vibrations also exhibits phonon linewidth. The integrated EPC parameter $\lambda$ for Mg$_2$RhH$_6$ is 2.62. The logarithmic average frequency ($\omega_{\log }$) can be derived from the A-D equation of 352.9 K. The $T_c$ is predicted to be 59 K. Similar to Mg$_2$RhH$_6$, the phonon linewidth in Mg$_2$IrH$_6$ is primarily contributed by high and intermediate-frequency H modes, as shown in Fig. 3(b). The high-frequency phonon modes in Mg$_2$IrH$_6$ show slight hardening compared to Mg$_2$RhH$_6$, where the frequency of $E_g$ modes increases from 0.149 to 0.189 eV. The EPC constant $\lambda$ for Mg$_2$IrH$_6$ is 2.56 with $\omega_{\log }$ of 391.4 K. The $T_c$ is predicted to be 65 K for Mg$_2$IrH$_6$. 

\subsubsection{Al$_2$MnH$_6$}

\begin{figure}[t]
	\includegraphics[width=1\linewidth]{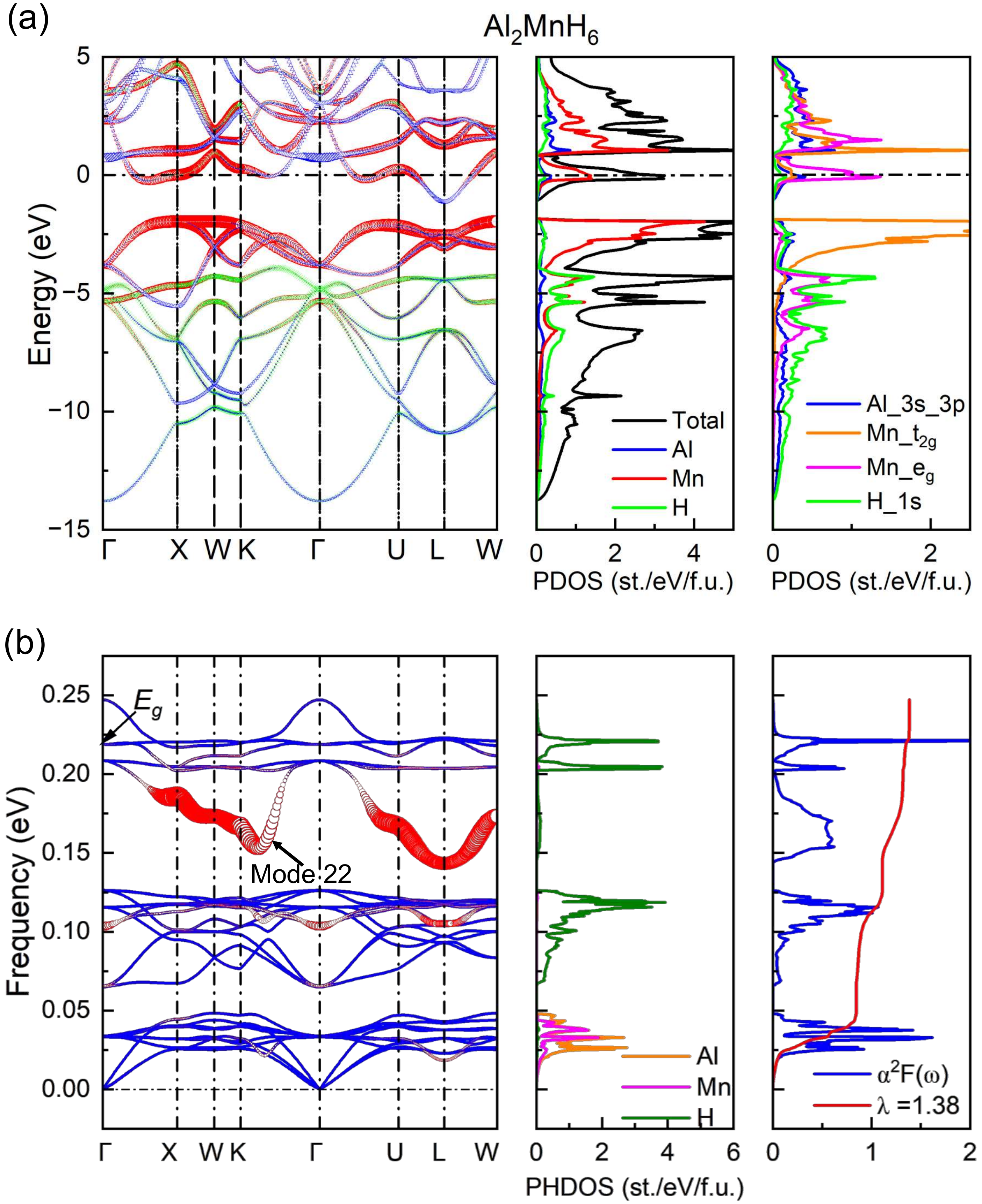}
	\caption{Electronic structure and electron-phonon calculations for Al$_2$MnH$_6$. (a) Electronic band structure and density of states projected into atoms and orbital. (b) The $\gamma_{q \nu}$-weighted (red circles) phonon spectrum, projected phonon density of states (PHDOS) and Eliashberg spectral function $\alpha^2 F(\omega)$.}
\end{figure}

Figure 4 (a) presents the electronic structure of Al$_2$MnH$_6$. Analogous to Mg$_2$RhH$_6$ and Mg$_2$IrH$_6$, a van Hove singularity is observed at $E_f$, contributed by flat bands along the high-symmetry path (2/3$\Gamma$-X-1/2W and 1/2$\Gamma$-U). The calculated total DOS is 3.25 eV$^{-1}$ per f.u, primarily contributed from transition metal (Mn) antibonding $e_g^*$ states mixed with H (0.21 eV$^{-1}$ per f.u) and Al states. The weaker Mn $t_{2g}$ nonbonding bands are located near 2.5 eV, exhibiting a sharp peak. The Mn $e_g$ bonding bands are at a lower energy level, showing strong hybridization with H-1$s$ bands over a broad energy range, approximately from -5 to -14 eV.

EPC calculations for Al$_2$MnH$_6$ are shown in Fig. 4(b). The phonon spectrum of Al$_2$MnH$_6$ is divided into three distinct regions. The phonon modes of Al and Mn atoms are located in the low-frequency range, from 0 to $\sim$ 0.048 eV. The H modes are in two regions, from $\sim$ 0.065 to $\sim$ 0.126 eV and $\sim$ 0.142 to $\sim$ 0.247 eV. Unlike Mg$_2$RhH$_6$ and Mg$_2$IrH$_6$, the doubly degenerated phonon modes $E_g$ at the $\Gamma$ point do not exhibit a strong phonon linewidth in Al$_2$MnH$_6$. Only the high-frequency mode 22 of the H atom contributes significantly to the phonon linewidth along the Brillouin zone boundary. The H modes at $\sim$ 0.102 eV along $\Gamma$ and L points also exhibit substantial phonon linewidth. The EPC constant $\lambda$ for Al$_2$MnH$_6$ is 1.38, lower than that of Mg$_2$RhH$_6$ and Mg$_2$IrH$_6$. However, the predicted $T_c$ of Al$_2$MnH$_6$ can reach up to 66 K, due to a high logarithmic average phonon frequency $\alpha^2 F(\omega)$ ($\sim$ 631.1 K).

\subsubsection{Li$_2$CuH$_6$}

\begin{figure}[t]
	\includegraphics[width=1\linewidth]{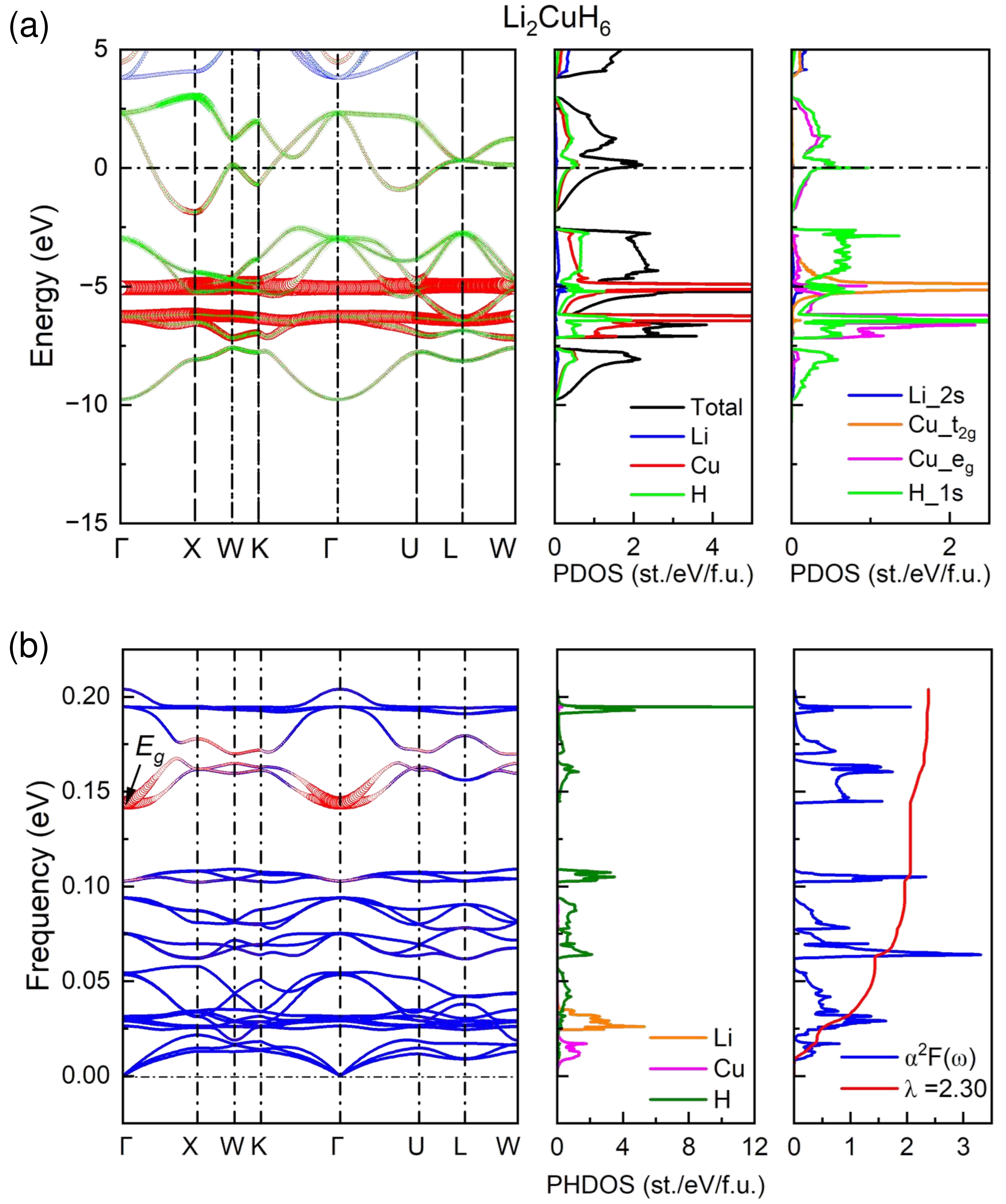}
	\caption{Electronic structure and electron-phonon calculations for Li$_2$CuH$_6$. (a) Electronic band structure and density of states projected into atoms and orbital. (b) The $\gamma_{q \nu}$-weighted (red circles) phonon spectrum, projected phonon density of states (PHDOS) and Eliashberg spectral function $\alpha^2 F(\omega)$.}
\end{figure}

Figure 5(a) displays the electronic structure of Li$_2$CuH$_6$. It also exhibits a van Hove singularity at $E_f$, with a total DOS of 1. 51 eV$^{-1}$ per f.u., smaller than that of Mg$_2$RhH$_6$, Mg$_2$IrH$_6$ and Al$_2$MnH$_6$. The contribution of Cu-$e_g^*$ antibonding states and H-1$s$ states (~0.40 eV$^{-1}$ per f.u.) is almost equal at $E_f$. In the valence band, the Cu-$t_{2g}$ nonbonding bands and $e_g$ bonding bands exhibit sharp peaks, resulting in distinct flat bands along the entire high-symmetry path.  

Figure 5(b) presents EPC calculations for Li$_2$CuH$_6$. Similar to Mg$_2$RhH$_6$ and Mg$_2$IrH$_6$, the doubly degenerated phonon modes $E_g$ at the $\Gamma$ point in Li$_2$CuH$_6$ significantly contribute to the phonon linewidth, with a lower frequency of 0.143 eV. The integrated EPC parameter $\lambda$ for Li$_2$CuH$_6$ is 2.30. The $\omega_{\log }$ derived from Eq. (8), is 500.1 K, leading to a predicted $T_c$ of 80 K.
\\

\section{Conclusions}
By employing the high-throughput screening method of zone-center electron-phonon interaction, we thoroughly investigate the superconductivity of cubic ternary hydrides X$_2$MH$_6$ (X = Li, Na, Mg, Al, K, Ca, Ga, Rb, Sr and In; M is $3d$, $4d$, and $5d$ transition metal) at ambient pressure. We identify 26 compounds demonstrating dynamic stability with high $\lambda_\Gamma$. Among these, several have been verified to be superconductors through our EPC calculations. Notably, four compounds (Mg$_2$RhH$_6$, Mg$_2$IrH$_6$, Al$_2$MnH$_6$ and Li$_2$CuH$_6$) have a $T_c$ above 50 K and a low energy above convex hull ($E_d< 80$ meV/atom), making them promising candidates for experimental synthesis. The $T_c$ of Li$_2$CuH$_6$ is 80 K, which is higher than the liquid-nitrogen temperature. The calculated electronic structures of these four compounds reveal that they all possess a sharp van Hove singularity at the Fermi level, accounting for their high $T_c$. Our study indicates a significant number of candidates displaying superconductivity in cubic ternary hydrides X$_2$MH$_6$ system at ambient pressure, which also demonstrates an effective strategy to explore conventional superconductors in ternary hydrides by taking into account distinctive structural motifs like MH$_6$ octahedron. This work is expected to reignite interest in investigating the new ternary hydrides with high $T_c$ at ambient pressure. \\

\section{Acknowledgments}
\noindent The work at Jimei University was supported by the Natural Science Foundation of Xiamen City (Grant No. 3502Z202372015) and the Research Foundation of Jimei University (Grant No. ZQ2023013). The work at Xiamen University was supported by the Natural Science Foundation of Xiamen (Grant No. 3502Z202371007) and the Fundamental Research Funds for the Central Universities (Grant No. 20720230014). V.A. was supported by the U.S. Department of Energy, Office of Basic Energy Sciences, Division of Materials Sciences and Engineering. Ames National Laboratory is operated for the U.S. Department of Energy by Iowa State University under Contract No. DE-AC02-07CH11358. K.-M.H. acknowledges support from National Science Foundation Award No. DMR2132666.\\


\begin{thebibliography}{}
\bibitem{1} A. A. Abrikosov, Sov. Phys. JETP, 14, 408 (1962).
\bibitem{2} N. W. Ashcroft, Phys Rev Lett, 21, 1748 (1968).
\bibitem{3} E. Gregoryanz, C. Ji, P. Dalladay-Simpson, B. Li, R. T. Howie, H. K. Mao, Matter Radiat Extrem, 5, 038101 (2020).
\bibitem{4} C.B. Satterthwaite, I. L. Toepke, Phys Rev Lett, 25, 741 (1970).
\bibitem{5} B. Stritzker, W. Buckel, Z. Phys. A: Hadrons Nucl, 257, 1 (1972).
\bibitem{6} B. Stritzker, Z. Phys., 268, 261 (1974).
\bibitem{7} D. Duan, Y. Liu, F. Tian, D. Li, X. Huang, Z. Zhao, H. Yu, B. Liu, W. Tian, T. Cui, Sci Rep, 4, 6968 (2014).
\bibitem{8} F. Peng, Y. Sun, C.J. Pickard, R. J. Needs, Q. Wu, Y. Ma, Phys Rev Lett, 119, 107001 (2017).
\bibitem{9} H. Liu, Naumov, II, R. Hoffmann, N. W. Ashcroft, R. J. Hemley, Proc Natl Acad Sci U S A, 114, 6990 (2017).
\bibitem{10} H. Wang, J. S. Tse, K. Tanaka, T. Iitaka, Y. M. Ma, Proc Natl Acad Sci U S A, 109, 6463 (2012).
\bibitem{11} H. Xie, Y. Yao, X. Feng, D. Duan, H. Song, Z. Zhang, S. Jiang, S. A.T. Redfern, V. Z. Kresin, C. J. Pickard, et al., Phys Rev Lett, 125, 217001 (2020).
\bibitem{12} B. B. Liu, W. W. Cui, J. M. Shi, L. Zhu, J. Chen, S. Y. Lin, R. M. Su, J. Y. Ma, K. Yang, M. L. Xu, et al., Physical Review B, 98, 174101 (2018).
\bibitem{13} A. M. Shipley, M. J. Hutcheon, R. J. Needs, C. J. Pickard, Phys Rev B, 104, 054501 (2021).
\bibitem{14} Z. P. Wu, Y. Sun, A. P. Durajski, F. Zheng, V. Antropov, K. M. Ho, S. Q. Wu, Physical Review Materials, 7, L101801 (2023).
\bibitem{15} L. L. Liu, F. Peng, P. Song, X. H. Liu, L. Y. Zhang, X. W. Huang, C. Y. Niu, C. Y. Liu, W. F. Zhang, Y. Jia, et al., Physical Review B, 107, L020504 (2023).
\bibitem{16} K. Gao, W. Cui, J. Shi, A. P. Durajski, J. Hao, S. Botti, M. A. L. Marques, Y. Li, Physical Review B, 109, 014501 (2024).
\bibitem{17} S. Q. Wu, M. Ji, C. Z. Wang, M. C. Nguyen, X. Zhao, K. Umemoto, R. M. Wentzcovitch, K. M. Ho, J Phys-Condens Mat, 26, 035402 (2014).
\bibitem{18} Y. Wang, J. Lv, L. Zhu, Y. Ma, Computer Physics Communications, 183, 2063 (2012).
\bibitem{19} J. Wang, H. Gao, Y. Han, C. Ding, S. Pan, Y. Wang, Q. Jia, H. T. Wang, D. Xing, J. Sun, Natl Sci Rev, 10, nwad128 (2023).
\bibitem{20} S. Saha, S. Di Cataldo, F. Giannessi, A. Cucciari, W. von der Linden, L. Boeri, Physical Review Materials, 7, 054806 (2023).
\bibitem{21} W. E. Pickett, Rev Mod Phys, 95, 021001 (2023).
\bibitem{22} A. P. Drozdov, M. I. Eremets, I. A. Troyan, V. Ksenofontov, S. I. Shylin, Nature, 525, 73 (2015).
\bibitem{23} A. P. Drozdov, P. P. Kong, V. S. Minkov, S. P. Besedin, M. A. Kuzovnikov, S. Mozaffari, L. Balicas, F. F. Balakirev, D. E. Graf, V. B. Prakapenka, et al., Nature, 569, 528 (2019).
\bibitem{24} M. Somayazulu, M. Ahart, A. K. Mishra, Z. M. Geballe, M. Baldini, Y. Meng, V. V. Struzhkin, R. J. Hemley, Phys Rev Lett, 122, 027001 (2019).
\bibitem{25} L. Ma, K. Wang, Y. Xie, X. Yang, Y. Wang, M. Zhou, H. Liu, X. Yu, Y. Zhao, H. Wang, et al., Phys Rev Lett, 128, 167001 (2022).
\bibitem{26} Z. W. Li, X. He, C. L. Zhang, X. C. Wang, S. J. Zhang, Y. T. Jia, S. M. Feng, K. Lu, J. F. Zhao, J. Zhang, et al., Nat Commun, 13, 2863 (2022).
\bibitem{27} P. Kong, V. S. Minkov, M. A. Kuzovnikov, A. P. Drozdov, S. P. Besedin, S. Mozaffari, L. Balicas, F. F. Balakirev, V. B. Prakapenka, S. Chariton, et al., Nat Commun, 12, 5075 (2021).
\bibitem{28} E. Snider, N. Dasenbrock-Gammon, R. McBride, X. Wang, N. Meyers, K. V. Lawler, E. Zurek, A. Salamat, R. P. Dias, Phys Rev Lett, 126, 117003 (2021).
\bibitem{29} Y. Y. Wang, K. Wang, Y. Sun, L. Ma, Y. C. Wang, B. Zou, G. T. Liu, M. Zhou, H.B. Wang, Chinese Phys B, 31, 106201 (2022).
\bibitem{30} J. A. Flores-Livas, L. Boeri, A. Sanna, G. Profeta, R. Arita, M. Eremets, Phys Rep, 856, 1 (2020).
\bibitem{31} L. Boeri, G. B. Bachelet, J Phys Condens Matter, 31, 234002 (2019).
\bibitem{32} R. Vocaturo, C. Tresca, G. Ghiringhelli, G. Profeta, J Appl Phys, 131, 033903 (2022).
\bibitem{33} Y. He, J. Lu, X.Q. Wang, J. J. Shi, Physical Review B, 108, 054515 (2023).
\bibitem{34} Y. He, J. J. Shi, Nano Lett, 23, 8126 (2023).
\bibitem{35} A. Sanna, T. F. T. Cerqueira, Y.-W. Fang, I. Errea, A. Ludwig, M. A. L. Marques, Prediction of Ambient Pressure Conventional Superconductivity above 80 K in Thermodynamically Stable Hydride Compounds, 2023, pp. arXiv:2310.06804.
\bibitem{36} T. F. T. Cerqueira, A. Sanna, M. A. L. Marques, Adv Mater, 36, 2307085 (2024).
\bibitem{37} K. Dolui, L. J. Conway, C. Heil, T. A. Strobel, R. Prasankumar, C. J. Pickard, Feasible route high temperature ambient-pressure hydride superconductivity, 2023, pp. arXiv:2310.07562 
\bibitem{38} S. S. S. Raman, D. J. Davidson, J. L. Bobet, O. N. Srivastava, J Alloy Compd, 333, 282 (2002).
\bibitem{39} Y. Sun, F. Zhang, C.-Z. Wang, K.-M. Ho, I. I. Mazin, V. Antropov, Phys Rev Mater, 6, 074801 (2022).
\bibitem{40} F. Zheng, Y. Sun, R. Wang, Y. Fang, F. Zhang, S. Wu, C.-Z. Wang, V. Antropov, K.-M. Ho, Physical Review B, 107, 014508 (2023).
\bibitem{41} F. Zheng, Y. Sun, R.H. Wang, Y. M. Fang, F. Zhang, S. Q. Wu, Q. B. Lin, C. Z. Wang, V. Antropov, K. M. Ho, Phys Chem Chem Phys, 25, 32594 (2023).
\bibitem{42} Y. Sun, Z. Zhang, A. P. Porter, K. Kovnir, K. M. Ho, V. Antropov, Npj Comput Mater, 9, 204 (2023).
\bibitem{43} R. Wang, Y. Sun, F. Zhang, F. Zheng, Y. Fang, S. Wu, H. Dong, C. Z. Wang, V. Antropov, K. M. Ho, Inorg Chem, 61, 18154 (2022).
\bibitem{61} S. Chen, Z. Wu, Z. Zhang, S.Q. Wu, K.-M. Ho, V. Antropov and Y. Sun, High-throughput screening for boride superconductors, arXiv:2310.06804 (2024).
\bibitem{44} G. Kresse, D. Joubert, Physical Review B, 59, 1758 (1999).
\bibitem{45} G. Kresse, J. Furthmuller, Physical Review B, 54, 11169 (1996).
\bibitem{46} G. Kresse, J. Furthmuller, Comp Mater Sci, 6, 15 (1996).
\bibitem{47} J. P. Perdew, K. Burke, M. Ernzerhof, Phys Rev Lett, 77, 3865 (1996).
\bibitem{48} H. J. Monkhorst, J. D. Pack, Physical Review B, 13, 5188 (1976).
\bibitem{49} A. Togo, F. Oba, I. Tanaka, Physical Review B, 78, 134106 (2008).
\bibitem{50} A. Togo, I. Tanaka, Scripta Mater, 108, 1 (2015).
\bibitem{51} P. Giannozzi, S. Baroni, N. Bonini, M. Calandra, R. Car, C. Cavazzoni, D. Ceresoli, G. L. Chiarotti, M. Cococcioni, I. Dabo, et al., J Phys-Condens Mat, 21, 395502 (2009).
\bibitem{52} P. Giannozzi, O. Andreussi, T. Brumme, O. Bunau, M. B. Nardelli, M. Calandra, R. Car, C. Cavazzoni, D. Ceresoli, M. Cococcioni, et al., J Phys-Condens Mat, 29, 465901 (2017).
\bibitem{53} S. Baroni, S. de Gironcoli, A. Dal Corso, P. Giannozzi, Rev Mod Phys, 73, 515 (2001).
\bibitem{54} M. J. van Setten, M. Giantomassi, E. Bousquet, M. J. Verstraete, D. R. Hamann, X. Gonze, G.M. Rignanese, Computer Physics Communications, 226, 39 (2018).
\bibitem{55} P. B. Allen, R. C. Dynes, Physical Review B, 12, 905 (1975).
\bibitem{56} P. B. Allen, Physical Review B, 6, 2577 (1972).
\bibitem{57} A. Jain, S. P. Ong, G. Hautier, W. Chen, W. D. Richards, S. Dacek, S. Cholia, D. Gunter, D. Skinner, G. Ceder, et al., Apl Mater, 1, 011002 (2013).
\bibitem{58} W. H. Sun, S. T. Dacek, S. P. Ong, G. Hautier, A. Jain, W. D. Richards, A. C. Gamst, K. A. Persson, G. Ceder, Science Advances, 2, e1600225 (2016).
\bibitem{59} K. Spektor, W. A. Crichton, S. Konar, S. Filippov, J. Klarbring, S. I. Simak, U. Haussermann, Inorg Chem, 57, 1614 (2018).
\bibitem{60} W. Bronger, S. Hasenberg, G. Auffermann, Zeitschrift f{\"u}r anorganische und allgemeine Chemie, 622, 1145 (2004).


\end{thebibliography}
\bibliographystyle{apsrev4-1}

\end{document}